\documentclass[aps, prb, reprint, showpacs]{revtex4-1}
\pdfoutput=1
\usepackage{amsmath}
\usepackage{graphicx}
\usepackage{braket}
\usepackage{siunitx}
\usepackage[caption=false]{subfig}
\usepackage{hyperref}

\begin{document}
\title{Nonequilibrium enhancement of high-temperature superconductivity in a 3D model of cuprates}
\date{\today}
\author{Zachary M. Raines}
\author{Valentin Stanev}
\affiliation{Joint Quantum Institute and Condensed Matter Theory Center, Department of Physics, University of Maryland, College Park, Maryland 20742-4111, USA}
\author{Victor M. Galitski}
\affiliation{Joint Quantum Institute and Condensed Matter Theory Center, Department of Physics, University of Maryland, College Park, Maryland 20742-4111, USA}
\affiliation{School of Physics, Monash University, Melbourne, Victoria 3800, Australia}
\begin{abstract}
Recent experiments in the cuprates have seen evidence of a transient superconducting state upon optical excitation polarized along the c-axis [R. Mankowsky et al., Nature \textbf{516}, 71 (2014)].
Motivated by these experiments we propose an extension of the single-layer $t-J-V$ model of cuprates to three dimensions in order to study the effects of inter-plane tunneling on the competition between superconductivity and bond density wave order. We find that an optical pump can suppress the charge order and simultaneously enhance superconductivity, due to the inherent competition between the two. We also provide an intuitive picture of the physical mechanism underlying this effect. Furthermore, based on a simple Floquet theory we estimate the magnitude of the enhancement.
\end{abstract}
\maketitle

\section{Introduction}
\label{sec:introduction}

Ever since the discovery of high-temperature superconductivity in cuprate materials\cite{Bednorz1986}, it has presented a fundamental challenge to our understanding of strongly interacting systems. Despite the decades of strenuous efforts these materials still remain a profound mystery. In the last several year, however, new experimental results have shone some light on one of the most puzzling phases of cuprates -- the pseudogap\cite{*[{For a review see e.g }] [{}] Keimer2014,*Taillefer2010a}. Clear evidence of some sort of charge ordering competing with superconductivity has emerged\cite{Chang2012,Fujita2014}.

Recently, in a series of exciting papers, it has been demonstrated that optical pumping can significantly enhance the superconductivity in several hole-doped cuprates \cite{Fausti2011,Kaiser2014,Hu2014,Nicoletti2014} (for a brief summary see Ref.~\onlinecite{Armitage2014}). In these works, a sample was subjected to a short pulse of mid-infrared light, polarized along the c-axis. Reflectivity measurements were then taken as a function of time delay, from which the frequency dependent conductivity was extracted. There is an increase in it that the authors identify with an inhomogeneous and transient enhancement of superconductivity. Furthermore, the nature of this enhancement seems to be different from the already well known effect due to quasiparticle photo-excitation\cite{Hu2014,eliashberg1970film,Ivlev1973,Chang1977,Robertson2009}. Several works have investigated these experiments and proposed an increase of inter-layer coupling as one of the dominant effects\cite{Mankowsky2014a,Hoppner2014a}.
However, there is another possible effect to consider, since resonant x-ray spectroscopy seems to suggest that the charge order is suppressed in the same region of the phase diagram where superconductivity is enhanced in the optical excitation experiments\cite{Forst2014}. A natural question to ask, then, is how coexisting superconductivity and charge order behave under such perturbation.

To address this problem we start from the $t-J-V$ model of the quasi two dimensional CuO$_2$ planes\cite{Kivelson1990,Dagotto1992}, which can naturally support the coexistence and competition between charge ordering and superconductivity\cite{Sau2013,Chowdhury2014,Allais2014}. Furthermore, we focus on the low energy physics of fermions near the so-called `hot spots', where the Fermi surface intersects the magnetic Brillouin zone\cite{Metlitski2010,Sachdev2013}. We extend the model by introducing an effective Hamiltonian, describing stacked planes, coupled by a c-axis tunneling term $t_z$. Then we investigate the phase diagram of this extended model by utilizing a Landau expansion of the free energy. Quite surprisingly, we observe a non-monotonic behavior of the critical temperatures of the two orders with increasing $t_z$, and we provide an intuitive physical explanation of this interesting feature. Finally, we consider different effects of the photo-excitations of the system, particularly focusing on the role of the apical oxygens, which are thought to play a key role in the experiments\cite{Kaiser2014,Hu2014,Mankowsky2014a}. We find that, quite generally, there is a parameter region where an optical pump can lead to a melting of the charge order and a corresponding enhancement in superconductivity, due to the competition between the two orders.


As a side note, let us add that the $t-J-V$ model and its variants tend to have as their leading instability a $(Q,Q)$ type ordering vector\cite{Sau2013,Chowdhury2014,Allais2014a}, with the $(Q, 0)$ ordering vector seen in experiment\cite{Fujita2014} as a sub-leading instability (with both orders having predominantly $d$-wave symmetry). While several extensions of the model have been proposed as a way to stabilize the experimentally observed order\cite{Allais2014,Chowdhury2014a,Thomson2014}, they introduce additional, and for our purposes unnecessary, complications. In order to keep the model simple, we restrict our attention to the instability at wavevector $(Q, Q)$. The physical content of our results lead us to expect that the qualitative behavior of the effects would be similar for the experimentally relevant $(Q, 0)$ charge order.

The structure of the paper is as follows. First, we provide a short description of the two-dimensional $t-J-V$ model and the associated Landau theory in Section~\ref{sec:2d_model}. In Section~\ref{sec:stack} we extend this model to include c-axis tunneling, and explore the effects of $t_z$ on the phase diagram. Time-dependent perturbation of $t_z$ is introduced in Section~\ref{sec:floquet}, and the high-frequency limit is studied. In Section~\ref{sec:conclusion} we summarise and discuss our findings.

\section{Two-dimensional Model}
\label{sec:2d_model}
In order to study the interplay of bond density wave (BDW) and superconducting orders we employ a 2D $t-J-V$ model of a CuO$_2$ plane\cite{Kivelson1990,Dagotto1992,Sau2013,Allais2014}. 
It provides a natural platform for exploring the general features of the interaction and the coupling between these two orders within a single copper oxide plane. The Hamiltonian is
\begin{equation}
    H = \sum_{i,j} t_{ij} c^\dag_{\sigma,i} c_{\sigma,j}
    + \frac{1}{2} \sum_{\langle i,j\rangle} J \vec{S}_i\cdot \vec{S}_j + \frac{1}{2} \sum_{\langle i, j\rangle} V n_i n_j,
    \label{eq:t-J-V}
\end{equation}
where $V$ and $J$ are nearest neighbor interactions, $n_i = \sum_\sigma c^\dag_{i,\sigma} c_{i,\sigma}$ is the charge density, and $\vec S_i = \frac{1}{2}\sum_\sigma c^\dag_{i,a}\vec{\sigma}_{ab} c_{i,b}$ is the site spin density, with $i$ and $j$ being site indices, and $\sigma$, $a$, and $b$ are spin indices. The term $t_{ij}$ contains nearest, next to nearest, and next to next to nearest hopping\cite{Parameters}. $V$ describes the nearest-neighbor tail of the Coulomb repulsion, which tends to suppress the $d$-wave superconductivity and enhance the BDW order.
$J$ is the usual nearest neighbor anti-ferromagnetic exchange interaction. 

Of course, the pure $t-J$ model has been extensively used in the studies of cuprates as an effective one-band description of the CuO$_2$ planes\cite{Lee2006}. It naturally leads to a $d$-wave superconductivity as its dominant instability. In order to have a region where superconductivity and BDW order coexist the model can be extended by the introduction of $V$, which suppresses superconductivity and boosts the charge order -- this is the rationale behind the $t-J-V$ model.

\begin{figure}[htpb]
    \centering
    \vspace{14pt}
    \includegraphics[width=0.8\linewidth]{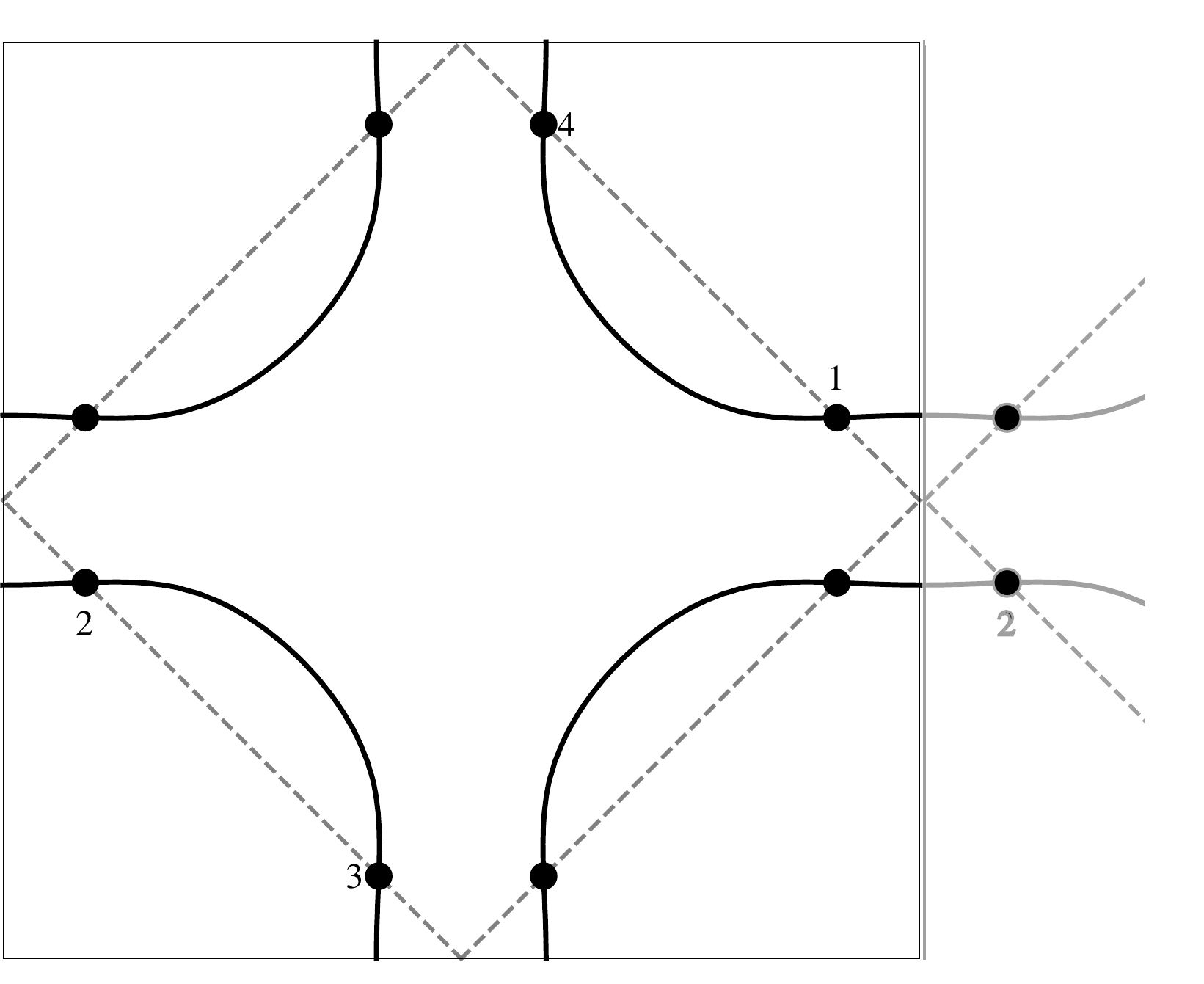}
    \caption{Hotspots exist where the Fermi surface intersects the magnetic brillioun zone boundary. The charge ordering vectors are given by the separations between hotspots 1 and 2 accross the Brillouin zone border and the rotated vector $(Q, -Q)$.\label{fig:hotspots}}
\end{figure}

We now construct a low-energy effective model by restricting our attention to fermions living within a limited region surrounding `hotspots' -- points where the Fermi surface intersects the magnetic Brillioun zone boundary, as depicted in Fig.~\ref{fig:hotspots} (such models have been recently introduced and used in a number of studies of charge order in cuprates\cite{Sau2013,Efetov2013,Wang2014a,Chowdhury2014}). These points are of special interest because the Fermi surface is nested with wave-vector $\vec K = (\pi, \pi)$, and given the importance of anti-ferromagnetic spin fluctuations to pairing in the cuprates\cite{Scalapino1995,Dahm2009,Sau2013}, we expect that the most relevant interactions will be those with exchanged momentum $\vec K$. Close to the hotspots we can replace the interactions $J_{\vec q}, V_{\vec q}$ with constants $J_{\vec K}, V_{\vec K}$, and restrict the two unconstrained fermion momenta $\vec{k} - \vec{k}' = \vec{q}$ to lie in hot regions separated by $\vec K$.

Furthermore, we only consider a charge ordering instability with a diagonal $(Q, Q)$ ordering vector and a d-wave form factor, as this was found to be the strongest mean-field instability\cite{Sau2013,Sachdev2013}. 
If we further enforce time reversal symmetry this allows us to concentrate our attention to four hotspots. Having restricted our fermions to live within a range $\Lambda$ of the hotspots we obtain an effective Hamiltonian
\begin{multline}
        H = \sum_{\vec k,i} \xi_{i,\vec k} c^\dag_{i,\vec k,\sigma}  c_{i,\vec k,\sigma}\\
    + g^{abcd} \sum_{\vec k,\vec p} \left[c^\dag_{1,\vec k,a}c_{2,\vec k,d} c^\dag_{4,\vec p,c} c_{3,\vec p,b}\right.\\
    \left. - c^\dag_{1,\vec k,a} c^\dag_{2,-\vec k,c} c_{4,-\vec p,d} c_{3,\vec p,b}\right],
\end{multline}
where
\[\sum_{\vec k} \dotsi \simeq \int_{-\Lambda}^\Lambda \int_{-\Lambda}^\Lambda \frac{d^2k}{(2\pi)^2} \dotsi,\]
the interaction is
\begin{equation}
    g^{abcd} = -\frac{1}{4}J_{\vec K} \vec{\sigma}_{ab} \cdot \vec{\sigma}_{cd} - V_{\vec K} \delta_{ab} \delta_{cd},
\end{equation}
$\vec k$ is now the deviation from the hotspot, $a-d$ are the electron spin indices, and $i$ is now a hotspot index.

At this point we undertake a mean field decomposition of the interaction (four-fermion) terms of the Hamiltonian simultaneously in the BDW and superconducting channels, where a $d$-wave form factor is assumed for both orders, by defining
\begin{equation}
\begin{aligned}
    i \sigma^y_{ab} \bar \Delta &=-&g_s&\sum_{\vec k} \braket{c^\dag_{1,\vec k,a} c^\dag_{2,-\vec k,b}} \\
    &= &g_s&\sum_{\vec k} \braket{c^\dag_{3,\vec k,a} c^\dag_{4,-\vec k,b}},
\end{aligned}
    \label{eq:delta}
\end{equation}
and
\begin{equation}
    \phi \delta_{ab} = -g_c\sum_{\vec k} \braket{c^\dag_{1,\vec k,a} c_{2,\vec k,b}} = g_c\sum_{\vec k} \braket{c^\dag_{3,\vec k,a} c_{4,\vec k,b}},
    \label{eq:phi}
\end{equation}
where
\begin{equation}
\begin{gathered}
g_s = \frac{1}{2}\sum_{a,b,c,d} g^{abcd}(i \sigma^y_{ac})(i\sigma^y_{db}) = 3J - 4V\\
g_c = \frac{1}{2}\sum_{a,b,c,d} g^{abcd}\delta_{ad}\delta_{bd} = 3J+4V,
\end{gathered}
\end{equation}
are the effective couplings in the superconducting and charge channels respectively, and $\braket{\cdots}$ indicates a thermal average. The order $\Delta$ is the projection of uniform d-wave superconductivity onto the hotspots, and $\phi$ describes a $d$-wave charge order, that lives on the bonds between copper sites, and has modulation vector $\vec Q$ given by the separation between hotspots. It is clear that $V$ enhances superconductivity and simultaneously suppresses charge order.
Because of the $d$-wave form factor of the order parameters, two of the remaining hotspots become redundant and at the mean field level the behavior of the system may be described by a $4\times4$ Hamiltonian in hotspot-Nambu space. Defining a Nambu spinor $\psi_{\vec{k}} = \left(c_{1,\vec{k},\uparrow}, c_{2,\vec{k},\uparrow}, c^\dag_{2, -\vec{k},\downarrow}, c^\dag_{1, -\vec{k}, \downarrow}\right)^T$ the Hamiltonian takes the form
\begin{equation}
    \mathcal{H} = \sum_{\vec k} \psi^\dag_{\vec k} \hat H_{\text{MF}}(\vec k) \psi_{\vec k} + \frac{2}{g_s} |\Delta|^2 + \frac{2}{g_c} |\phi|^2,
    \label{eq:Hmf}
\end{equation}
where
\begin{equation}
   \hat H_{\text{MF}}(\vec k) =
    \begin{bmatrix}
        \xi_1(\vec k) & \bar \phi & \Delta & 0\\
        \phi & \xi_2(\vec k) & 0 & \Delta\\
        \bar \Delta & 0 & -\xi_1(\vec k) & - \bar \phi\\
         0 &\bar \Delta & - \phi & -\xi_2(\vec k)
    \end{bmatrix},
\end{equation}
and $\Delta$ and $\phi$ are the superconductivity and BDW order parameters respectively.

Both order parameters are generally complex numbers, which we can write as $\Delta = |\Delta| e^{i \theta_\Delta}$ and $\phi = |\phi| e^{i \theta_\phi}$. However, since we can always remove the complex phases (at the mean-field level) via a gauge transformation\footnote{The ability to gauge away the phase degrees of freedom requires that we be considering only superconductivity and a single charge order and applies only to the hotspot model at the mean field level.} $c_1 \to c_1 e^{-i (\theta_\Delta - \theta_\phi)/2}$, $c_2 \to c_2 e^{-i (\theta_\phi + \theta_\Delta)/2}$, we will consider only real and non-negative values for $\phi$ and $\Delta$ in our analysis.

From \autoref{eq:Hmf} we can readily derive a Landau free energy for the $\Delta$ and $\phi$ orders. Evaluating $f=f_{MF} + \frac{1}{N} \braket{H - H_{MF}}_{MF}$, using the above decoupling and expanding to fourth order in the order parameters, we obtain
\begin{multline}
    f = \left[\frac{2}{g_c} - \Pi_{BDW}\right]\phi^2 + \left[\frac{2}{g_s} - \Pi_{SC}\right] \Delta^2\\
    + u_{BDW} \phi^4 + u_{SC} \Delta^4 + w \phi^2 \Delta^2,
    \label{eq:2dlandau}
\end{multline}
with $w, u_{BDW}, u_{SC} > 0$. The exact expressions for the coefficients are given in \autoref{tab:coeff2d}. It is in fact possible to solve the mean field problem exactly via a sequence of Bogoliubov transformations. This method breaks down, however, once we introduce c-axis hopping, and so, in anticipation of this extension of the model, we have chosen instead to work with a Landau expansion, which will carry over to the more complicated case.

\begin{table}[htbp]
    \begin{tabular}{lc}
        Term&Expression\\
        \hline
        $\Pi_{SC}$&$\int_{\vec k} \frac{1 - 2n(\xi_1, T)}{\xi_1}$\\
        $\Pi_{BDW}$ & $2 \int_{\vec k} \frac{n(\xi_{2}, T) - n(\xi_1, T)}{\xi_1 - \xi_{2}}$\\
        $u_{SC}$ & $\int_{\vec k} \frac{1}{2\xi_1^2}\left[ n'(\xi, T) +  \frac{1 - 2n(\xi, T)}{2\xi_1}\right]$\\
        $u_{BDW}$ & $\int_{\vec k} \frac{1}{{(\xi_1-\xi_{2})}^2}\left[ n'(\xi_1, T) + n'(\xi_{2}, T) + 2 \frac{n(\xi_{2}, T) - n(\xi_1, T)}{\xi_1 - \xi_{2}}\right]$\\
        $w$ & $2\int_{\vec k} \frac{1}{\xi_1 (\xi_1 - \xi_{2}) } \left[n'(\xi_1, T) + \frac{1-2 n(\xi_1, T)}{2 \xi_1} \right]$
    \end{tabular}
    \caption{Microscopic expressions for coefficients of the Landau theory. $n$ is the Fermi function and $n'$ the derivative of the Fermi function with respect to energy.\label{tab:coeff2d}}
\end{table}

\begin{figure}[htpb]
    \centering
    \includegraphics[width=0.9\linewidth]{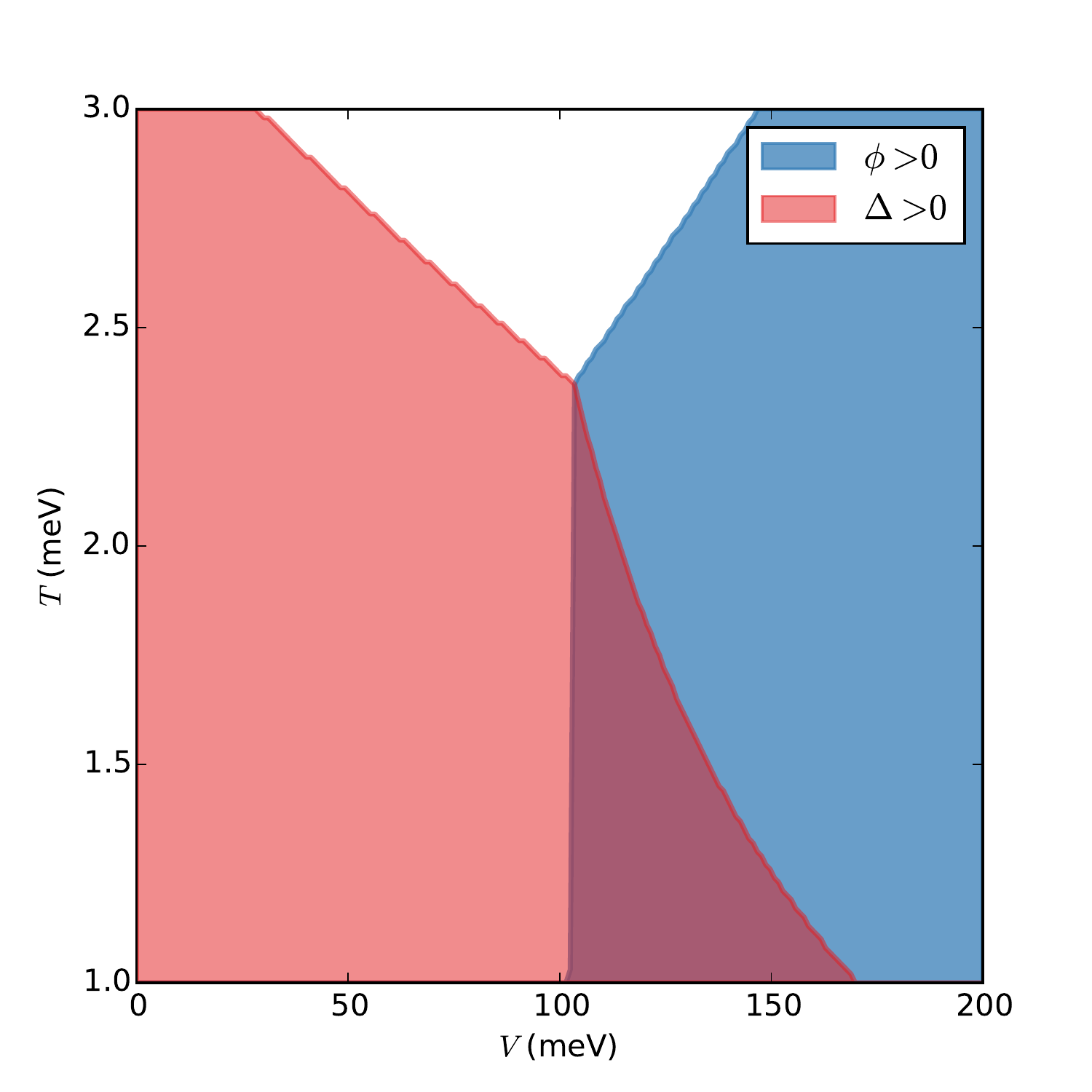}
    \caption{(Color online) Phase diagram of the 2D model\cite{Parameters}, for fixed $J$. The regions are from left to right, SC, BDW + SC, and BDW, with the white region at the top being the normal phase. Note that the temperature does not extend down to $T=0$ as the Landau theory is not valid in this limit.\label{fig:PD0L}}
\end{figure}

In what follows we hold $J$ fixed and use $V$ to adjust the splitting between the superconducting and BDW instabilities. 
Doing so, we obtain a phase diagram such as that in \autoref{fig:PD0L}. In it we see the following easy-to-understand behavior: when $V$ is zero or very small the $d$ wave superconductivity is the only relevant instability of the model. With $V$ increasing, superconductivity is suppressed and eventually BDW appears as the leading order. Notably, in the region where these instabilities are comparable there is a (rather narrow) coexistence phase. These features are consistent with previous studies of the model\cite{Sau2013}.

 We expect this phase diagram to be strictly valid only close to $T_c$ \emph{and} in the region in which the critical temperatures of both orders are comparable.  However, comparing the Landau expansion with the exact numerical mean-field solution shows that even for intermediate temperatures there is only a small, purely quantitative correction to the shape of the coexistence region.

\section{Extension to stacked planes}
\label{sec:stack}
Motivated by the experiments which have used optical excitation polarized along the c-axis to create transient states of enhanced superconductivity\cite{Hu2014,Kaiser2014,Nicoletti2014}, we seek to extend the purely two-dimensional model from the previous section to include coupling between planes.
The simplest model that encapsulates this behavior can be represented by the following Hamiltonian:
\begin{equation}
H_{3D} = \sum_l H_l + H_\text{T}.
\label{eq:H_3D}
\end{equation}
Here $l$ is a layer index, and $H_l$ is the single-layer Hamiltonian considered in \autoref{eq:t-J-V}.
There are $N_z$ copies of those, which are coupled via a c-axis tunneling term
\begin{equation}
H_\text{T} = \sum_{\vec{k}_\parallel, k_z} c^\dag_{\vec{k}_\parallel, k_z\sigma} t_z(\vec{k}_\parallel, k_z) c_{\vec{k}_\parallel, k_z, \sigma},
\label{eq:c-tun}
\end{equation}
where c-axis momentum $k_z$ is conjugate to the plane index $l$.

In what follows we will consider and compare three different forms for the tunneling $t_z$ in Eq.~\ref{eq:c-tun}. 
The exact expressions for each tunneling type are presented in \autoref{tab:tun}. Type A tunneling  (nearest neighbor hopping along the c-axis) we introduce mainly for its simplicity, type B comes from a one-band tight binding fit to band structure calculations for LSCO\cite{Markiewicz2005a}, and type C was proposed as an approximate tunneling form for several families of cuprate superconductor\cite{Xiang2000}. Despite the significant differences between these tunneling forms, it turns out that the effects we obtain are not specific to any of them, but are in all cases qualitatively similar.

\begin{table}[htpb]
    \centering
    \begin{tabular}{lc}
        Type & c-axis tunneling\\
        \hline
        A& $-2t_z \cos k_z$\\
        B\cite{Markiewicz2005a}&$-2t_z \cos\left(\frac{k_z}{2}\right) \left(\cos k_x -\cos k_y\right)^2 \cos \left(\frac{k_x}{2}\right) \cos \left(\frac{k_y}{2}\right)$\\
        C\cite{Xiang2000}&$-2t_z\cos (k_z) (\cos k_x - \cos k_y)^2$
    \end{tabular}
    \caption{c-axis tunneling elements used in the calculation\label{tab:tun}}
\end{table}
We now retrace the same steps as in \autoref{sec:2d_model}. The derivation proceeds similarly, but there are some subtleties that need to be considered first.

\begin{figure}[htpb]
    \centering
    \includegraphics[width=0.7\linewidth]{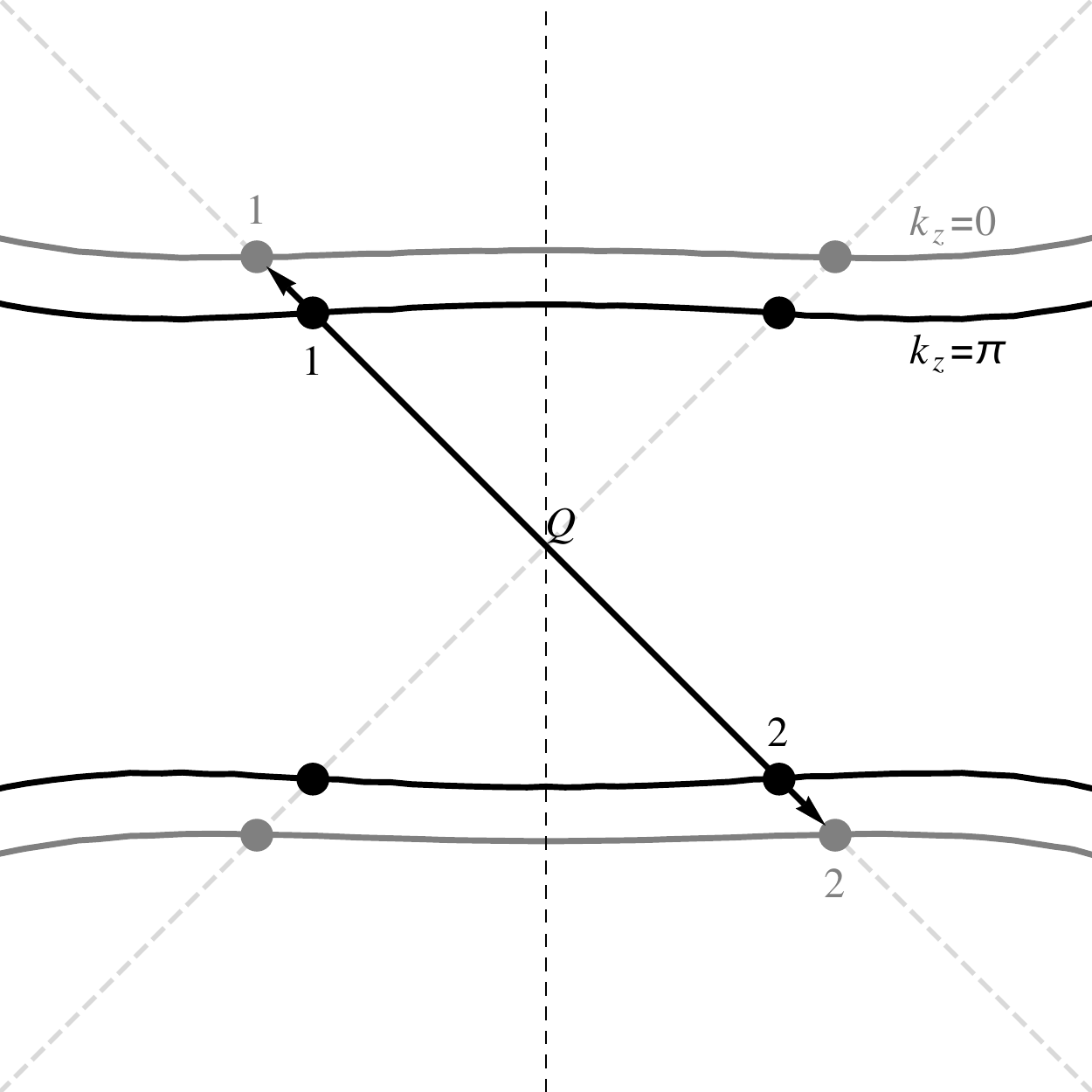}
    \caption{c-axis hopping causes a warping of the Fermi surface which causes the hotspots to move as a function of $k_z$ and leads to a destruction of $\vec Q$ nesting away from $k_z = 0$.\label{fig:destnesting}}
\end{figure}

Due to the model containing solely in-plane interactions, we only need to consider pairing of quasi-particles within the same plane. As a consequence of this, the order parameters do not depend on $k_z$ and the vector $\vec Q$ which separates fermions contributing to pairing in the charge channel cannot change with $k_z$. Because of this restriction, we find that while at $k_z=0$ superconductivity and BDW pair the same points in the 2D Brillioun zone, this is no longer true for $k_z\neq 0$, $t_z\neq 0$. The particle and hole being paired in the BDW channel must remain separated by $\vec{Q} = (Q, Q)$ even when the Fermi surface is no longer nested with this vector, as can be seen in \autoref{fig:destnesting}. Superconductivity on the other hand continues to pair $\vec k$ and $-\vec k$ for all $k_z$. As a consequence, one has to be careful to define the hot regions properly in this case. There are several ways one might do so, but fortunately, as shown in the appendix, they all lead to the same qualitative behavior.

That being the case, we implement the following procedure. At each $k_z$ there is a region centered on where the 2D Fermi surface intersects the 2D magnetic Brillioun zone. $\phi$ now pairs quasi-particles separated by the fixed charge ordering vector and is only non-zero when the momenta of both fall within a hot region. With this procedure in place, the free energy takes the same form as \autoref{eq:2dlandau} but with the coefficients now being the three dimensional integrals shown in \autoref{tab:coeff}. Here, we have made the simplifying assumptions that $\Delta$ and $\phi$ are not modulated along the $c$-axis. As can readily be seen, the expressions in \autoref{tab:coeff} reduce to those in \autoref{tab:coeff2d} in the limit of no c-axis hopping.

\begin{table*}[htbp]
    \begin{tabular}{lc}
        Term&Expression\\
        \hline
        $\Pi_{SC}$&$\int_{\vec k} \frac{1 - 2n(\xi_1, T)}{\xi_1}$\\
        $\Pi_{BDW}$ & $2 \int'_{\vec k} \frac{n(\xi_{1-Q}, T) - n(\xi_1, T)}{\xi_1 - \xi_{1-Q}}$\\
        $u_{SC}$ & $\int_{\vec k} \frac{1}{2\xi^2}\left[ n'(\xi, T) +  \frac{1 - 2n(\xi, T)}{2\xi}\right]$\\
        $u_{BDW}$ & $\int'_{\vec k} \frac{1}{{(\xi_1-\xi_{1-Q})}^2}\left[ n'(\xi_1, T) + n'(\xi_{1-Q}, T) + 2 \frac{n(\xi_{1-Q}, T) - n(\xi_1, T)}{\xi_1 - \xi_{1-Q}}\right]$\\
        $w$ & $2\int'_{\vec k} \frac{1}{\xi_1 (\xi_1 - \xi_{1-Q}) } \left[n'(\xi_1, T) + \frac{1-2 n(\xi_1, T)}{2 \xi_1} \right]$
    \end{tabular}
    \caption{Microscopic expressions for the coefficients of the Landau theory. $\xi_{1-Q}$ is the energy at a point at the point $\vec{k}' = \vec{k} + \vec{k}_\text{HS} - \vec{Q}$. Primed integration indicates a restriction of the integral to regions where both $\vec{k}$ and $\vec{k'}$ lie within a hot region. The $k_z$ integration is from $-\pi$ to $\pi$.\label{tab:coeff}}
\end{table*}

\begin{figure}[htpb]
    \includegraphics[width=0.95\linewidth]{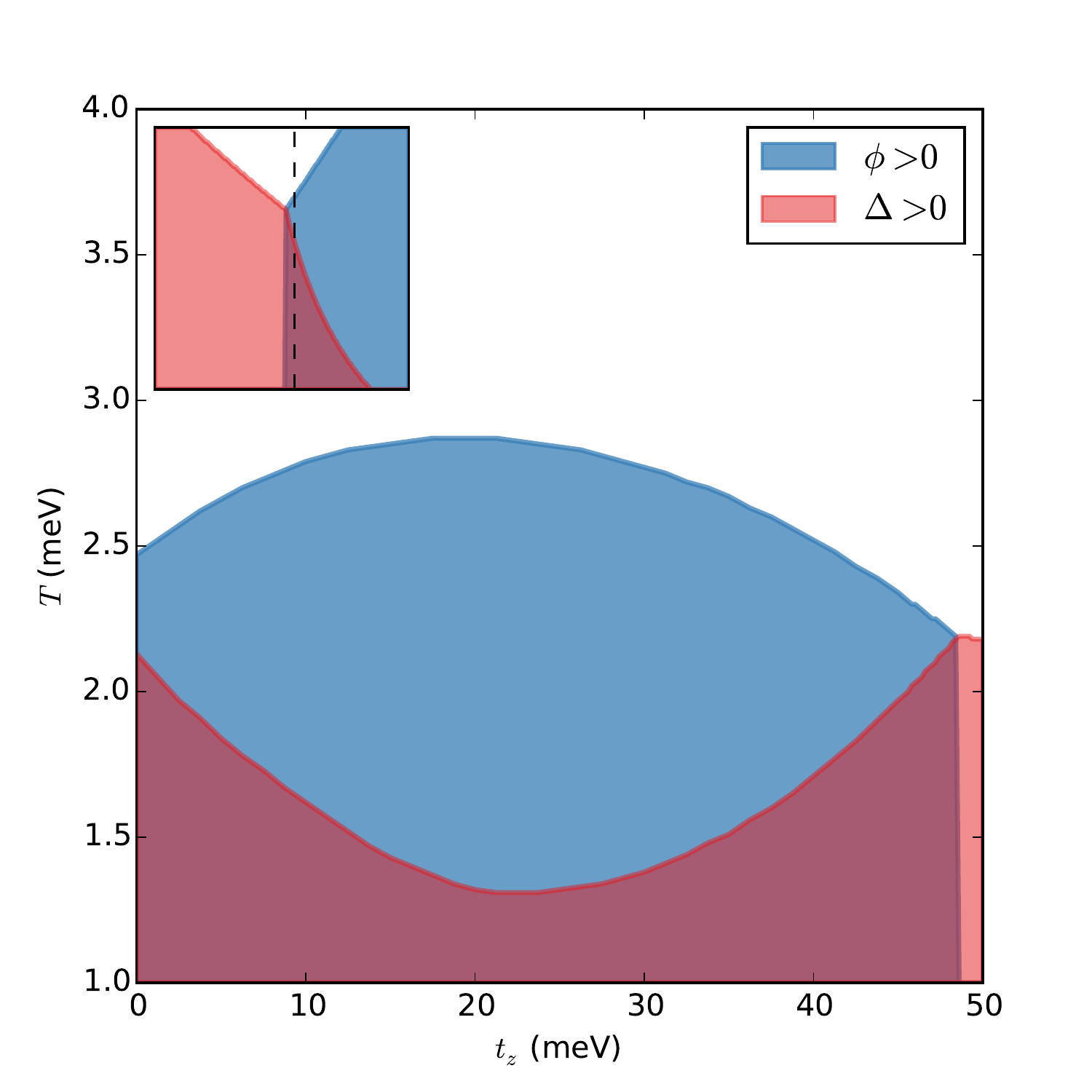}
    \caption{(Color online) Phase diagram of a system of stacked planes at a fixed coupling strength for tunneling type B in \protect\autoref{tab:tun}. Types A and C exhibit qualitatively similar behavior, but with different scales on the $t_z$ axis. The inset indicates where the starting point at $t_z=0$ lies in $V-T$ plane of \protect\autoref{fig:PD0L} ($V=\SI{110}{meV}$).\label{fig:tzphase}}
\end{figure}

Using each of the three tunneling forms, we calculated the state which minimized the free energy for a range of $T$, $V$ and $t_z$. The $(Q,Q)$ order remains the leading charge instability at the quadratic level, so we again only decouple in this channel and the d-wave superconducting channel. The phase diagram for tunneling type B is presented on Fig.~\ref{fig:tzphase} (the other two types lead to qualitatively similar diagrams). We start (for $t_z=0$) with the coexistence case in which BDW is the leading instability. As we can see, for small but finite $t_z$ the charge order is further enhanced at the expense of superconductivity. However, once $t_z$ becomes sufficiently large, the tendency reverses, and superconductivity is boosted by the increase of three-dimensionality, until it becomes the leading instability.

\begin{figure}[htpb]
    \centering
    \subfloat[Quadratic Susceptibilities]{\includegraphics[width=0.8\linewidth]{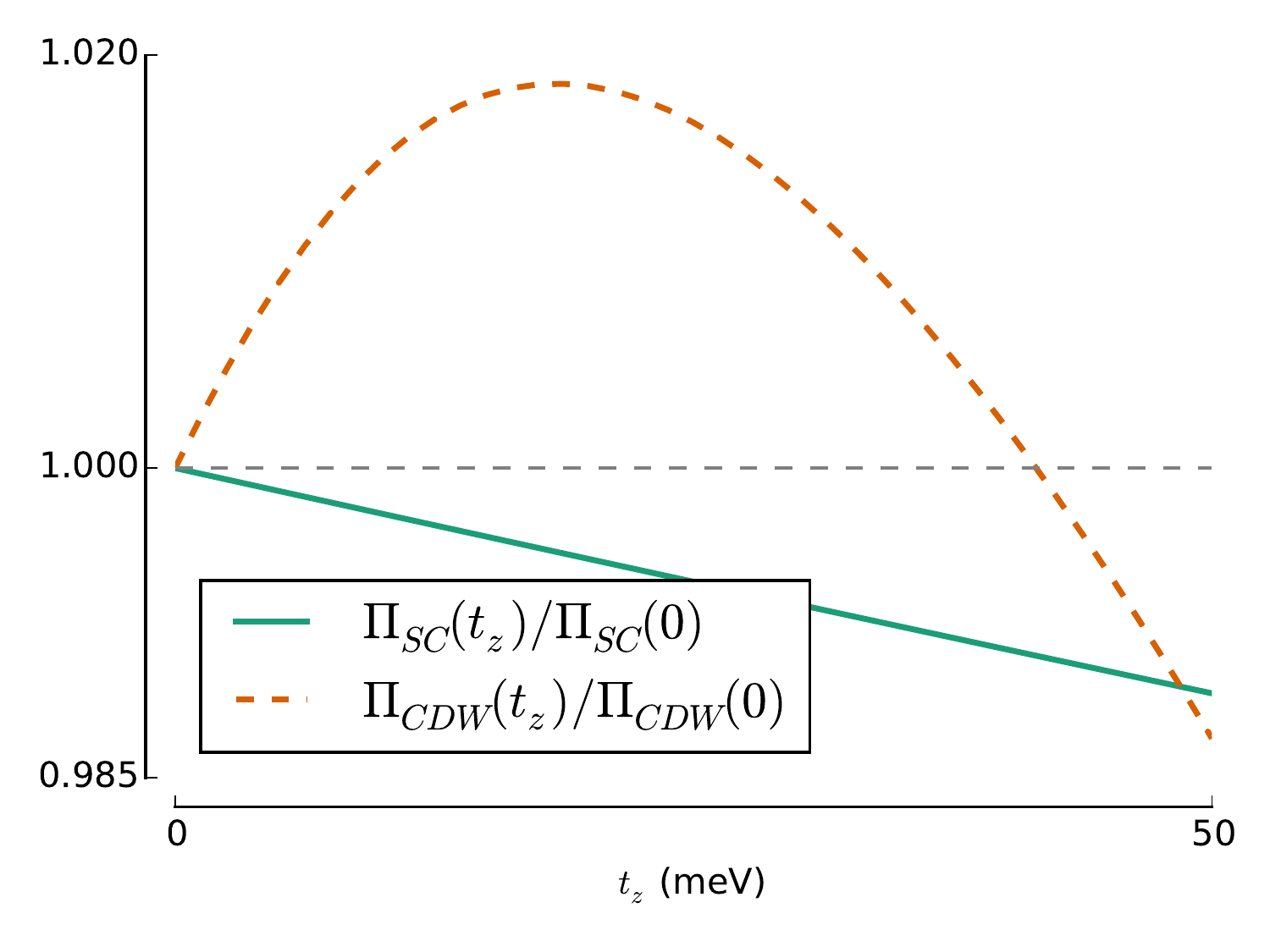}}\\
    \subfloat[Quartic Terms]{\includegraphics[width=0.8\linewidth]{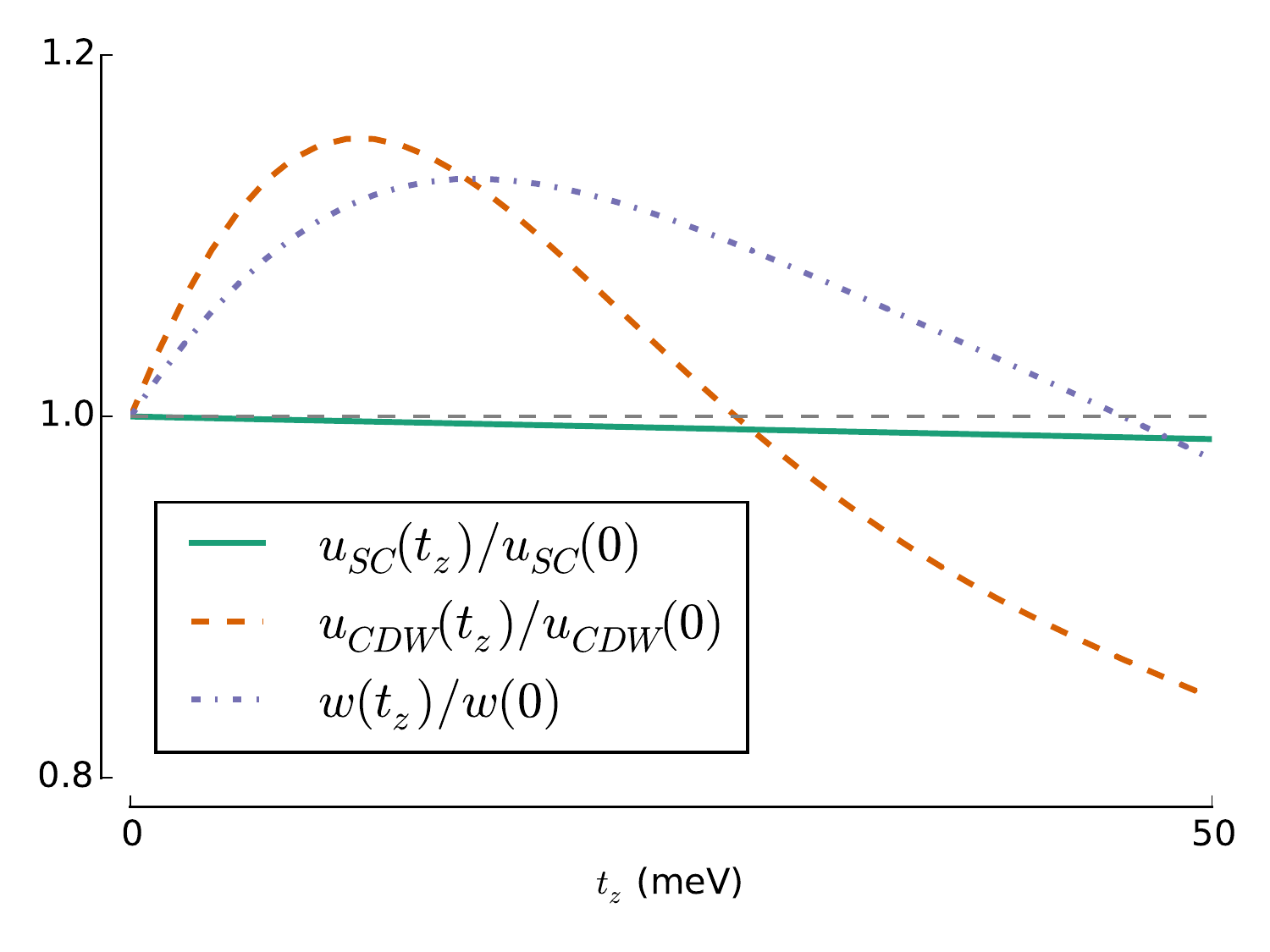}}
    \caption{(Color online) Quadratic susceptibilities and quartic coefficients of the Landau free energy as a function of $t_z$, scaled by their value at $t_z = 0$, for type B tunneling (at fixed temperature). Types A and C again exhibit similar behavior. Note that there is no significance to the crossing of lines for different coefficients, as the leading instability will be determined by $g\Pi$.\label{fig:quadsusc}}
\end{figure}

To understand this peculiar shape of the phase diagram, we look at the quadratic coefficients in the free energy. The behavior of the superconducting coefficients, $\Pi_{SC}$ and $u_{SC}$, can be explained entirely by the dependence of the density of states and bandwidth, $\rho(\xi, t_z)$ and $W(t_z)$. As can be seen in \autoref{fig:quadsusc}, these lead to only a small effect on the superconducting susceptibility. Thus, the features visible in \autoref{fig:tzphase} are mostly determined by the behavior of the charge instability and its indirect effect on superconductivity through the biquadratic term in the free energy.

The non-monotonic behavior of the charge susceptibility $\Pi_{BDW}$ can be understood as a competition of two effects. First, it is well known within this model that in the absence of Fermi surface curvature, the diagonal BDW order exhibits a BCS type transition characterized by a logarithmic divergence. Curvature however cuts off the divergence of the logarithm and weakens the charge instability\cite{Sachdev2013,Sau2013,Moor2014}. Specifically, we can write the dispersions as
\begin{gather}
    \xi_1 = \xi_+ + \xi_-\\
    \xi_2 = \xi_+ - \xi_-
\end{gather}
In that case the charge susceptibility becomes
\begin{equation}
    \Pi_{BDW} = \int_{\vec k} \frac{\sinh\frac{\xi_-}{2T}\cosh\frac{\xi_-}{2T}}{\xi_-[\sinh^2\frac{\xi_+}{2T} + \cosh^2\frac{\xi_-}{2T}]}
\end{equation}
in the 2D limit. One can see clearly that in the limit that $\xi_+ \to 0$, the absence of curvature, the charge (particle-hole) instability becomes as strong as the superconducting (particle-particle) one, and for any non-zero $\xi_+$ the charge susceptibility is weaker compared to that of the BCS case.

As the c-axis hopping $t_z$ increases the 2D curvature decreases. This is because the $k_z=0$ component for all tunneling types is of the form $-t_z\eta(\vec k_\parallel) c^\dag_{\vec k_\parallel} c_{\vec k_\parallel},$ with $\eta > 0$. Near the hotspots this acts as an effective upward shift in the chemical potential, which moves the hotspots such that $\xi_+$ is decreased relative to $\xi_-$. As a result, for increasing $t_z$, the $k_z=0$ Fermi surfaces changes in such a way as to effectively enhance the charge instability. Fundamentally, this is a consequence of the effect of the tunneling term on the properties of the Fermi surface of each plane.

In opposition to the aforementioned effect, increasing $t_z$ will lead to a progressive destruction of nesting away from $k_z=0$, as depicted in \autoref{fig:destnesting}. 
For $t_z=0$, the Fermi surface is nested at the hotspots for all $k_z$. However, as $t_z$ increases the Fermi surface warps more with $k_z$, decreasing the portion of the phase space for which there is approximate nesting, and thus weakening the charge susceptibility.

For small $t_z$ the warping of the Fermi surface is small, and so the 2D decrease of curvature is nearly the same for a wide range of $k_z$ values, leading to an overall strengthening of the charge instability. However, as $t_z$ increases Fermi surface warping along the c-axis becomes more pronounced. As a consequence, the available phase space for charge ordering is substantially reduced, while at the same time the significance of the decreased curvature is lessened away from $k_z=0$, together eventually causing a weakening of the charge ordering instability. The non-monotonic shape of the phase diagram can consequently be understood as demonstrating a crossover between regimes in which the 2D and the 3D effects of $t_z$ dominate.

Interestingly, the value of $t_z$ for which superconductivity reaches a minimum is of similar magnitude to the strength of c-axis hopping for various families of cuprates obtained as a fit to band structure\cite{Markiewicz2005a}. Given this, if we imagine a system where $t_{z}$ is near or greater than the point of minimum superconducting $T_c$ in \autoref{fig:tzphase}, then enhancements of $t_z$ will generally lead to suppression of charge order and enhancement of superconductivity.


It should be noted that the effects demonstrated here are obtain within the grand canonical ensemble at fixed chemical potential. In anticipation of \autoref{sec:floquet} the experimental program we envision comprises placing a sample in contact with its environment, which functions as a particle bath maintaining $\mu$, and then applying perturbations that will lead to a change of $t_z$ in the effective Hamiltonian. This, however, means that the volume enclosed by the Fermi surface is not constant and therefore average particle number is not conserved. If one were instead to consider a system where such a constraint were important this might change the behavior of the phase diagram at small $t_z$, where the effects are largely governed by the position of the $k_z=0$ Fermi surface. However, larger $t_z$ should still lead to Fermi surface warping and thus the enhancement of superconductivity in that regime should remain. Therefore, while the 2D effects that are important predominantly at small $t_z$ could be washed away, the suppression of charge order (via destruction of nesting), and thus the enhancement of superconductivity, appears much more robust.

\section{Enhancement of superconductivity via periodic modulation}
\label{sec:floquet}


Very recently, it has been argued that in YBCO the c-axis vibrations, induced by the external field, can change the equilibrium lattice structure\cite{Mankowsky2014a}. The result is a transient shift of the CuO$_2$ planes -- the intra-bilayer  distance increases, while the inter-bilayer one decreases. In our model this would lead to effectively bringing the layers closer. Intuitively, it is clear that this should lead to enhancement of the inter-plane coupling $t_z$. In this the role of the apical oxygen seems quite significant; previous experimental works have found that there is a range of dopings for which the inter-layer hopping exhibits a roughly exponential dependence on the hole-doping\cite{Nyhus1994} while the bond length between the plane copper and the apical oxygen exhibits a roughly linear dependence on doping through the same region\cite{Jorgensen1990}. A natural interpretation is that the hopping exhibits an exponential dependence on an effective barrier width $d$:
\begin{equation}
    t_z = A e^{-\alpha d},
    \label{eq:tindhop}
\end{equation}
where $d$ is approximately linear in the distance between the plane and the apical oxygen\cite{Nyhus1994,Honma2010}. Thus, decreasing the interlayer distance leads to an \emph{exponential} increase of the c-axis tunneling.

There is a second, more subtle, way in which driving the apical oxygens can enhance the interlayer tunneling. Let us consider a vibration of these ions \emph{without} change of their equilibrium positions.
Then, in line with the above reasoning, we can model the effect of harmonic oscillations of the apical oxygen on $t_z$ as oscillations of the effective barrier width, leading to a time dependent hopping element
\begin{equation}
    t_z(t) = A e^{-\alpha d(t)} =  t_{z0} \exp [-\alpha d_1 \cos(\Omega t)],
    \label{eq:timehop}
\end{equation}
where $d(t) = d_0 + d_1 \cos(\Omega t)$, and $t_{z0}= A e^{-\alpha d_0}$.

Let us consider the high frequency limit. Then, we expect that the quasiparticles will see an effective time averaged Hamiltonian.
While in experiment the frequencies are not extremely high, we consider this limit as a particularly simple case, from which we may extract relevant qualitative trends. More formally, we can obtain a Floquet Hamiltonian related to the time dependent hoppings. The stroboscopic dynamics of the system will be governed by this Floquet Hamiltonian, which can be obtained as series in $1/\Omega$ via a Magnus expansion of the time evolution operator\cite{Bukov2014}. For a high frequency oscillation, we keep only the first term of this expansion, which is just the time-dependent Hamiltonian averaged over one period. In this case the Floquet Hamiltonian is the original Hamiltonian (\autoref{eq:H_3D}), but with the modification
\begin{equation}
    t_{z} \to \langle t_z(t)\rangle = \Omega \int_0^\frac{2 \pi}{\Omega} \frac{dt}{2\pi} t_z(t) = t_{z0} I_0(\alpha d_1),
\end{equation}
where $I_0$ is the modified Bessel function of the first kind. $I_0$ is bounded below by $1$, and increases monotonically with the magnitude of $d_1$. Therefore, within this approximation, any oscillation of the apical oxygens unavoidably leads to enhancement of the effective c-axis tunneling $t_z$. This can be easily understood  from the exponential dependence of the tunneling on the apical oxygen position: in the tail of the exponent, a stronger enhancement is obtained from decreasing the argument than the suppression when increasing it at $\pi/\Omega$ time later. Thus, the tunneling amplitude is, on average, enhanced. Note that this observation is rather general, and likely applicable well beyond the region of validity of the high-frequency approximation used above.

By using data from Refs.~\onlinecite{Nyhus1994,Jorgensen1990} we can obtain an estimate to the magnitude of both the oscillatory enhancement and the slower, quasi-static, effect\footnote{As discussed in Ref.~\onlinecite{Nyhus1994}, we may relate their measurements of the integrated c- spectral weight $N_\text{eff}^c$ to the c-axis tunneling strength via the c-axis plasma frequency. If we take the exponential behavior of this quantity to be dominated by the c-axis tunneling with form $t_z \propto e^{-\alpha d(x)}$, with $d$ the Cu(1)-O(4) (in-plane copper to apical oxygen) bond length, we may write $N_\text{eff}^c \sim A e^{-2\alpha d(x)}$, where $A$ is effectively a constant. The data in Ref.~\onlinecite{Nyhus1994} is for $N_\text{eff}^c$ as a function of doping, but Ref.~\onlinecite{Jorgensen1990} provides data showing a quasilinear dependence of the Cu(1)-O(4) bond length on doping. Thus, we model $d(x) = d_0 - \eta x$ with $x$ the hole doping. We may now obtain a rough estimate of $\eta\approx\SI{0.075}{\angstrom}$ and $\alpha \approx \SI{31}{\angstrom^{-1}}$ by approximating the data points presented in figures in Refs.~\onlinecite{Nyhus1994,Jorgensen1990} in order to reproduce the observed data.}.
Assuming an oscillation distance of $d_1 \sim \SI{2.2}{pm}$, as is observed in the experiments discussed in Ref.~\onlinecite{Mankowsky2014a}, we find approximately a $10-15\%$ enhancement of $t_z$ in the steady state average. This in turn may lead to a few percent up to around a $15\%$ enhancement in the superconducting $T_c$ depending where in the non-monotonic structure of Fig.~\ref{fig:tzphase} the sample is before perturbation. A stronger effect is the shift of the equilibrium position of the apical oxygen. A quasi-static shift of the apical oxygen position by $\SI{2.4}{pm}$ (again see Ref.~\onlinecite{Mankowsky2014a}) can enhance $T_c$ by up to about $60\%$. These effects provide a qualitative picture of a possible mechanism underlying the explanation given in Ref.~\onlinecite{Mankowsky2014a}.

\section{Discussion and Conclusion}
\label{sec:conclusion}

The primary motivation for this investigation came from the recent experiments on transient enhancement of superconductivity in the cuprates via mid-infrared optical excitations\cite{Mankowsky2014a,Kaiser2014,Hu2014,Nicoletti2014}. To model these experiments we considered an extension of the $t-J-V$ model of cuprates to three dimensions, and the effects of this three-dimensionality on the competition between superconductivity and bond density wave orders. We showed that for our extended model, increased inter-plane tunneling leads to a suppression of charge ordering, and a coinciding enhancement of superconductivity due to the inherent competition between the two orders. The evolution of charge order takes place in two steps, corresponding to the regions where 2D and 3D effects of increased interlayer coupling, respectively dominate. The primary effect of interest is that the charge instability is sensitive to the c-axis curvature of the Fermi surface, which destroys nesting at the charge ordering vector. This effect is generic across several tunneling forms proposed for various cuprate materials.

These results provide a physical picture explaining the enhancement of superconductivity by the decrease of the inter-bilayer distance, caused by optical excitation. We further showed that periodic oscillations of the apical oxygens (identified in the experiments as important) can also lead to an effective increase of the inter-layer coupling. Both these effects indirectly promote superconductivity via a suppression of the competing charge ordering. We believe that the mechanism presented here could play a significant part in the observed enhancement of superconductivity and could be useful in pursuing new ways to raise $T_c$. Other  mechanisms have been proposed with regard to these experiments: the suppression of phase fluctuations\cite{Hoppner2014a,Robertson2011} as well as the usual enhancement of superconductivity due to microwave stimulation\cite{eliashberg1970film} could certainly play a complementary role.

At the end, let us also note that this work considers only the mean field behavior of such a system. As is well known, fluctuations play an extremely important role in the superconducting transition of cuprates\cite{Emery1995,Corson1999,Xu2000}. Nevertheless, we expect that the effects discussed here on a mean field level will remain important in a more complete description of the system (entering through the relevant energy scales, for example). The mean field theory presented in this paper is only the first necessary step in the study of these effects, and including fluctuations is an important direction for future work.


\begin{acknowledgments}
We are grateful to J.D. Sau for enlightening discussions. 
This work was supported by DOE-BES DESC0001911 and Simons Foundation.
\end{acknowledgments}

\appendix*

\section{Low energy model away from \texorpdfstring{$k_z =0$}{kz=0}}
\label{sec:choices}
\begin{figure}[htpb]
    \centering
    \includegraphics[width=0.85\linewidth]{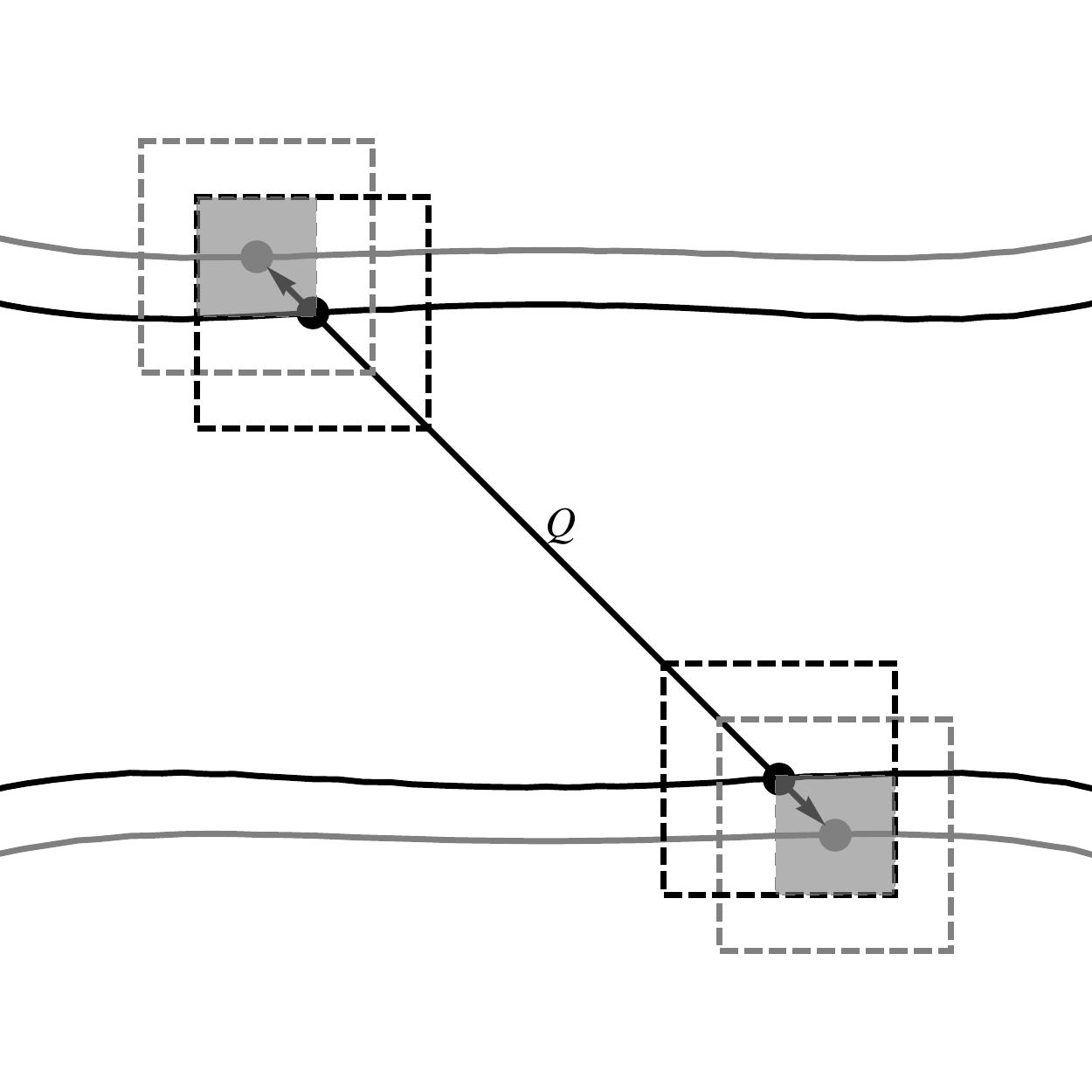}
    \caption{Different choices of hot regions for charge integrals. The gray dashed lines denotes the \emph{unmoving} hot regions. The black dashed lines represent the \emph{moving} hot regions, where one must be careful to avoid double counting. Finally the gray-filled areas represent the \emph{truncated} hot regions.\label{fig:choices}}
\end{figure}

As discussed in \autoref{sec:stack} there are subtleties associated with defining the hotspot model for stacked planes with inter-plane hopping. Specifically, one has to define the integrals involving charge order such that the separation between paired particles remains constant with $k_z$, despite the fact that the Fermi surface changes shape. We considered three different methods of handling this issue. Each is depicted in \autoref{fig:choices}. In all cases the coefficients $\Pi_{SC}$ and $u_{SC}$ are unchanged, so we only need to decide how to implement the other three: $\Pi_{BDW}$, $u_{BDW}$ and $w$.

\begin{figure}[htpb]
    \centering
    \includegraphics[width=0.95\linewidth]{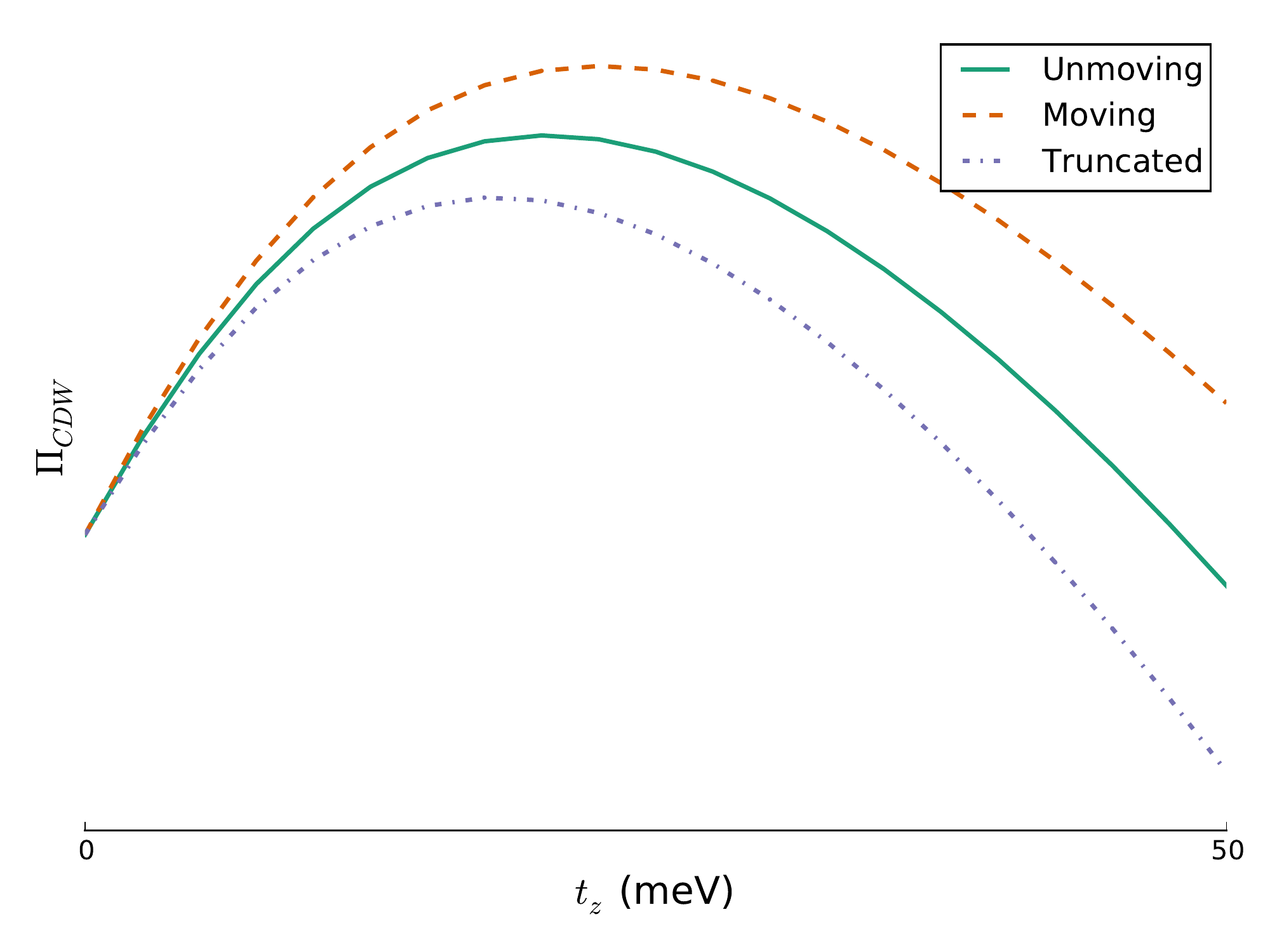}
    \caption{(Color online) Quadratic susceptibilities as a function of c-axis hopping for different choices of integration region in integrals involving charge order. The qualitative behavior is the same for all three schemes.\label{fig:hotspotchoice}}
\end{figure}

The first is to define two different hot regions: one remains bound to the Fermi surface and is associated with superconductivity, the other corresponds to the $k_z = 0$ hot regions for all $k_z$ and is associated with charge order. All integrals involving charge order are done over the unmoving hotspots. We label this procedure the \emph{unmoving} approximation.

Another approach is to perform the charge order integrations over the same hot regions as the superconducting terms, but enforce that pairing occur between particles separated by the fixed ordering vector $\vec Q$. We label this the \emph{moving} approximation.

A final approach, and the one we used in this work, is to take a similar tact as the moving approximation, but to further restrict the integration such that both of the paired particles lie in a hot region. We will call this the \emph{truncated} approximation. If we are committed to the notion that only fermions near the hotspots are important, this seems the most natural of the approximations, as we are then only counting the fermions that live within the true hot regions. The coefficients $\Pi_{BDW}$, $u_{BDW}$, and $w$ are shown for a range of $t_z$ in \autoref{fig:hotspotchoice}. As can be seen, while the results differ numerically, the qualitative behavior is the same. Since the hotspot model is itself a qualitative model, we chose to adopt the \emph{truncated} approximation as it seemed the most in line with the spirit of the model.

While these issues make the 3D model slightly more complicated, the qualitative behavior of the system appears to be fairly robust to their method of resolution. This suggests that the extended model, like its simpler progenitor, can be used as a tool to uncover general physical mechanisms underlying the complicated behavior of various 
 lattice models.

\bibliography{references,footnotes}
\end{document}